\pdfoutput=1

\documentclass{ws-ijbc}
\usepackage{graphicx}
\usepackage{multicol}
\usepackage{hyperref}

\hypersetup{pdftitle = The title of my PDF, pdfauthor = My name, pdfsubject= The subject, pdfkeywords = keyword1 keyword2 keyword3}

\hypersetup{colorlinks = true, linkcolor = blue, anchorcolor = red, citecolor = blue, filecolor = red, pagecolor = red, urlcolor = blue}

\newcommand{\mathz}{\ooalign{$z$\cr\hfil\rule[.5ex]{.2em}{.06ex}\hfil\cr}}

\begin{document}

\catchline{}{}{}{}{} % Publisher's Area please ignore

\markboth{Zotos et al.}{Basins of convergence in the circular Sitnikov four-body problem with non-spherical primaries}

\title{Basins of convergence in the circular Sitnikov four-body problem with non-spherical primaries}

\author{Euaggelos E. Zotos}

\address{Department of Physics, School of Science,
Aristotle University of Thessaloniki, \\
GR-541 24, Thessaloniki, Greece \\
Corresponding author's email: {evzotos@physics.auth.gr}}

\author{Satyendra Kumar Satya}

\address{Department of Mathematics, LNJ College,
Madhubani, Bihar, India}

\author{Rajiv Aggarwal}

\address{Department of Mathematics, Sri Aurobindo College,
University of Delhi, Delhi, India}

\author{Md Sanam Suraj}

\address{Department of Mathematics, Sri Aurobindo College,
University of Delhi, Delhi, India}

\maketitle

\begin{history}
\received{Received March 22, 2018}
\end{history}

\begin{abstract}
The Newton-Raphson basins of convergence, related to the equilibrium points, in the Sitnikov four-body problem with non-spherical primaries are numerically investigated. We monitor the parametric evolution of the positions of the roots, as a function of the oblateness coefficient. The classical Newton-Raphson optimal method is used for revealing the basins of convergence, by classifying dense grids of initial conditions in several types of two-dimensional planes. We perform a systematic and thorough analysis in an attempt to understand how the oblateness coefficient affects the geometry as well as the basin entropy of the convergence regions. The convergence areas are related with the required number of iterations and also with the corresponding probability distributions.
\end{abstract}


%\begin{multicols}{2}

\section{Introduction}
\label{intro}

The Sitnikov problem always refers to the special formulation of the restricted three-body problem, which describes the vertical motion of an infinitesimal mass $m$, along $z$-axis, perpendicular to the plane of the motion of two equally massed primaries, moving in circular or elliptic orbits with their common barycentre. The MacMillan problem \cite{McM11} corresponds to the case in which the primaries move in circular orbits. It was \cite{P07} who described the dynamical model in which the two primaries move in circular orbit around their common barycenter, which paved the path for the beginning of the Sitnikov problem.

For several decades, the Sitnikov problem remains a fascinating research topic, with a plethora of open topics, such as the periodic orbits via Poincar\'{e} maps (e.g., \cite{CL00}), the manifold of families of three-dimensional periodic orbits (e.g., \cite{Per07}), the periodic orbits in the case with prolate primaries (e.g., \cite{DKMP12}), the periodic orbits in the photogravitational Sitnikov three-body problem with oblateness (e.g., \cite{KPP08}), and various other aspects (e.g., \cite{C99,D93,F03,H92,H09,HL05,JP97,JE01,PK12,RGH15,SBD07}).

The natural extension of the Sitnikov restricted three-body problem is the Sitnikov restricted four-body problem which describes the motion of a test particle moving along the vertical $z$-axis, under the mutual gravitational attraction of three primaries of equal masses, moving in circular or elliptic orbits around their common barycenter. \cite{SPB08} discussed the periodic orbits and bifurcation in the restricted four-body problem in the Sitnikov sense. Furthermore, the stability of the vertical motion, in the Sitnikov sense, and its bifurcation in the $N$-body problem, have been revealed by \cite{BP09}. It was unveiled that there exists only one interval of stable vertical solution for every $N\geq 4$-body problem, which increases in size with the increase of the number of the primary bodies.

\cite{PA13a} investigated the Sitnikov four-body problem, by considering all the primaries as oblate spheroids which are symmetrical in all other respects. In their study, they revealed the relation between the side of equilateral triangle and the oblateness of the primaries to maintain the equilateral triangle configuration. In addition, they found only one stability region and twelve critical periodic orbits, from which new three-dimensional families of symmetric periodic orbits bifurcate and the stability interval increases with the increase of the oblateness parameter. Some other related studies on the Sitnikov four-body problem are described in \cite{PA13b,SH11,SH13,SH14}.

The study of the influence of the various perturbing parameters on the domain of the basins of convergence, associated with the libration points, using the Newton-Raphson iterative scheme unveils some of the most intrinsic properties of the dynamical system. Some of the pioneer works on the field of Newton-Raphson basins of convergence are the following: \cite{D10,Z17c} (for the Hill problem with oblateness and radiation), \cite{Z16} (for the restricted three-body problem with oblateness and radiation), \cite{Z17b} (for the pseudo-Newtonian restricted three-body problem), and \cite{BP11,KK14,SAA17,SAP17,Z17a} (for the restricted four-body problem with various types of perturbations).

The present paper is a result of the above-mentioned ideas which inspired us to introduce the oblateness of the primaries in the circular Sitnikov four-body problem. The main aim is to determine the influence of the oblateness parameter on the geometry as well as on the shape of the convergence domain, by using the Newton-Raphson iterative scheme. The layout of the article is as follows: the most important properties of the dynamical system are presented in Section \ref{mod}. The parametric evolution of the position of the equilibrium points is investigated in Section \ref{eqpts}. The following Section contains the main numerical results, regarding the evolution of the Newton-Raphson basins of convergence, while in Section \ref{pebe} we monitor the evolution of the basin entropy of the complex plane. Our paper ends with Section \ref{conc}, where we emphasize the main conclusions of this work.

\section{Properties of the mathematical model}
\label{mod}

Three primary bodies $P_i$, $i = 1, 2, 3$, with equal masses $m_i = m = 1/3$, are situated at the vertices of an equilateral triangle, while we consider a dimensionless, rotating, barycentric rotating system of coordinates $Oxyz$. The line passing through the center of the primary $P_1$ and center of the mass of the equilateral triangle is taken as the $x-$axis, while the line perpendicular to the $(x,y)$ plane of motion is taken as the vertical $z-$axis. Furthermore, we assume that the shape of the primaries is not spherically symmetric but it resembles a spheroid. Therefore, for each primary we introduce the corresponding oblateness coefficient $A_i$, $i = 1, 2, 3$. The centers of the three primaries are located at $(x_i, y_i, z_i)$ where
\begin{align}
x_1 &= \frac{a\sqrt{3}}{3}, \ \ \ y_1 = 0, \ \ \ z_1 = 0, \nonumber\\
x_2 &= - \frac{x_1}{2}, \ \ \ y_2 = \frac{a}{2}, \ \ \ z_2 = 0, \nonumber\\
x_3 &= - \frac{x_1}{2}, \ \ \ y_3 = - y_2, \ \ \ z_3 = 0,
\label{cents}
\end{align}
while $a = 1 + A$.

According to \cite{PA13a} the time-independent effective potential function of the circular restricted four-body problem with spheroid primaries is
\begin{equation}
\Omega(x,y,z) = \sum_{i=1}^{3}\frac{m_i}{r_i}\left(1 + \frac{A_i}{2 {r_i}^2} - \frac{3 A_i {z^2}}{2 {r_i}^4}\right) + \frac{1}{2}\left(x^2 + y^2 \right),
\label{pot}
\end{equation}
where
\begin{equation}
r_i = \sqrt{(x - x_i)^2 + (y - y_i)^2 + (z - z_i)^2}, \ \ \ i = 1, 2, 3,
\label{dist}
\end{equation}
are the distances of the fourth body from the respective primaries.

The equations of motion describing the dynamics of the fourth body (with negligible mass $m$), moving under the mutual gravitational attraction of the three primaries read as
\begin{equation}
\ddot{x} - 2\dot{y} = \frac{\partial{\Omega}}{\partial{x}}, \ \ \
\ddot{y} + 2\dot{x} = \frac{\partial{\Omega}}{\partial{y}}, \ \ \
\ddot{z} = \frac{\partial{\Omega}}{\partial{z}}.
\label{eqmot}
\end{equation}

\begin{figure}[!t]
\centering
\resizebox{0.5\hsize}{!}{\includegraphics{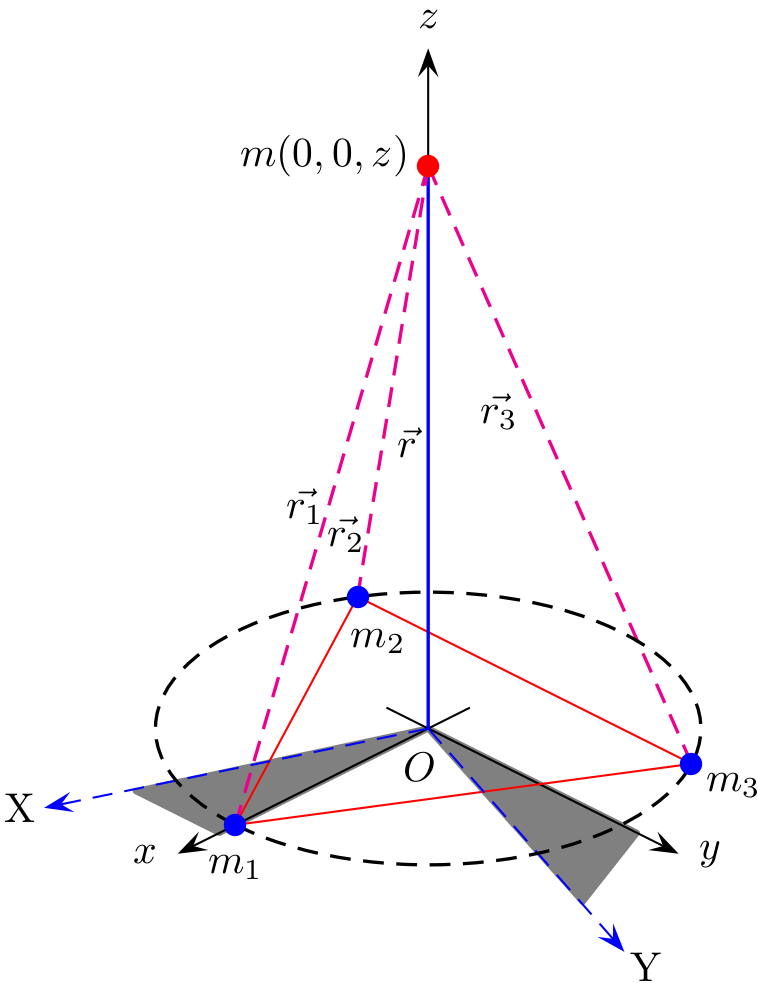}}
\caption{The configuration of the Sitnikov four-body problem, where three equally massed primary bodies $(m_1 = m_2 = m_3 = 1/3)$ move on symmetric circular orbits, on the same $(x,y)$ plane. The test particle of negligible mass $m$ is moving along the straight line which is perpendicular to the orbital plane of the primaries. (Color figure online).}
\label{conf}
\end{figure}

The above mentioned system of differential equations admits only one known integral of motion (also known as the Jacobi integral), which is described by the following Hamiltonian function
\begin{equation}
J(x,y,z,\dot{x},\dot{y},\dot{z}) = 2\Omega(x,y,z) - \left(\dot{x}^2 + \dot{y}^2 + \dot{z}^2 \right) = C,
\end{equation}
where $\dot{x}, \dot{y},$ and $\dot{z}$ represent the velocities, while the numerical value of the Jacobian constant is shown by $C$ and it is conserved.

If we set in Eq. (\ref{pot}) $m_i= 1/3$, $x = y = 0$, and $A_i = A$ then the same equation reduces to
\begin{equation}
\Omega(z) = \frac{1}{r} + \frac{A}{2r^3} - \frac{3 A z^2}{2r^5},
\label{potz}
\end{equation}
where $r = \sqrt{k^2 + z^2}$ with $k = \left(a\sqrt{3}\right)/3$. Eq. (\ref{potz}) is the potential function of the circular Sitnikov problem of four bodies and describes the motion of the fourth body (with mass $m$) which oscillates along the $z$-axis i.e. along the straight line perpendicular to the orbital $(x,y)$ plane of the primaries. In Fig. \ref{conf}, we present the configuration of the circular Sitnikov four-body problem.

Therefore the equation describing the motion of the fourth body along the vertical $z$-axis is
\begin{equation}
\ddot{z} = - \frac{z}{r^3} - \frac{9 A z}{2r^5} + \frac{15 A z^3}{2r^7},
\label{eqmotz}
\end{equation}
while the corresponding Jacobi integral, for the case of the vertical motion, reduces to
\begin{equation}
J(z,\dot{z}) = 2\Omega(z) - \dot{z}^2 = C_z.
\label{hamz}
\end{equation}

\section{Parametric variation of the equilibrium points}
\label{eqpts}

Following the approach successfully used in \cite{DKMP12} (see Section 3), from now on the $z$ coordinate is considered as a complex variable and it is denoted by $\mathz$.

The location of the positions of the equilibrium points can be obtained by setting the right hand side of  Eq.(\ref{eqmotz}) equal to zero which leads to
\begin{equation}
f(\mathz;A) = - \frac{\mathz}{r^3} - \frac{9 A \mathz}{2r^5} + \frac{15 A \mathz^3}{2r^7} = 0,
\label{fza0}
\end{equation}
which after simple calculations it reduces to
\begin{align}
&\mathz \left(18\mathz^4 + 6\left(A - 2\right)\left(2A - 1\right)\mathz^2 + a^2 \left(A\left(2A + 31\right) + 2\right)\right) \nonumber\\
&= 0.
\label{fza}
\end{align}

Equation (\ref{fza}) reveals that the root $z=0$ is always present, regardless the value of $A$ of the oblateness coefficient of the primaries. This root is directly associated with the inner collinear libration point $L_1$ of the circular restricted four-body problem. The left hand side of the Eq. (\ref{fza}) is a fifth order polynomial which leads to the fact that there are four additional roots, $z_i$, $i = 1, ..., 4$, given by
\begin{equation}
z_i = \pm \frac{\sqrt{-2A^2 + 5A - 2 \pm 3 \sqrt{-A \left(10A^2 + 11A +10\right)}}}{\sqrt{6}}.
\label{rts}
\end{equation}

Evidently, the nature of the roots strongly depends on the numerical value of  the oblateness coefficients $A$. Our analysis reveals that
\begin{itemize}
  \item When $A < A_1$ four pure imaginary roots exits, along the $\mathz = 0$ root.
  \item When $A = A_1$ two pure imaginary roots exist, along the $\mathz = 0$ root.
  \item When $A_1 < A < A_2$  two real and two pure imaginary roots exist, along the $\mathz = 0$ root.
  \item When $A = A_2$, only two pure imaginary roots exist.
  \item When $A_2 < A < A_3$ two real and two pure imaginary roots exist, along the $\mathz = 0$ root.
  \item When $A = A_3$  two pure imaginary roots exist, along the $\mathz = 0$ root.
  \item When $A_3 < A < A_4$  four pure imaginary roots exits, along the $\mathz = 0$ root.
  \item When $A = A_4$ only the $\mathz = 0$ root exists.
  \item When $A > A_4$ four complex roots exist, along the $\mathz = 0$ root.
\end{itemize}
The values
\begin{align}
A_1 &= \frac{\left(-31 - 3\sqrt{105}\right)}{4}, \nonumber\\
A_2 &= -1, \nonumber\\
A_3 &= \frac{\left(-31 + 3\sqrt{105}\right)}{4}, \nonumber\\
A_4 &= 0,
\label{crits}
\end{align}
are in fact critical values of the oblateness coefficient, since they determine the change on the nature of the four roots.

\begin{figure*}[!t]
\centering
\resizebox{\hsize}{!}{\includegraphics{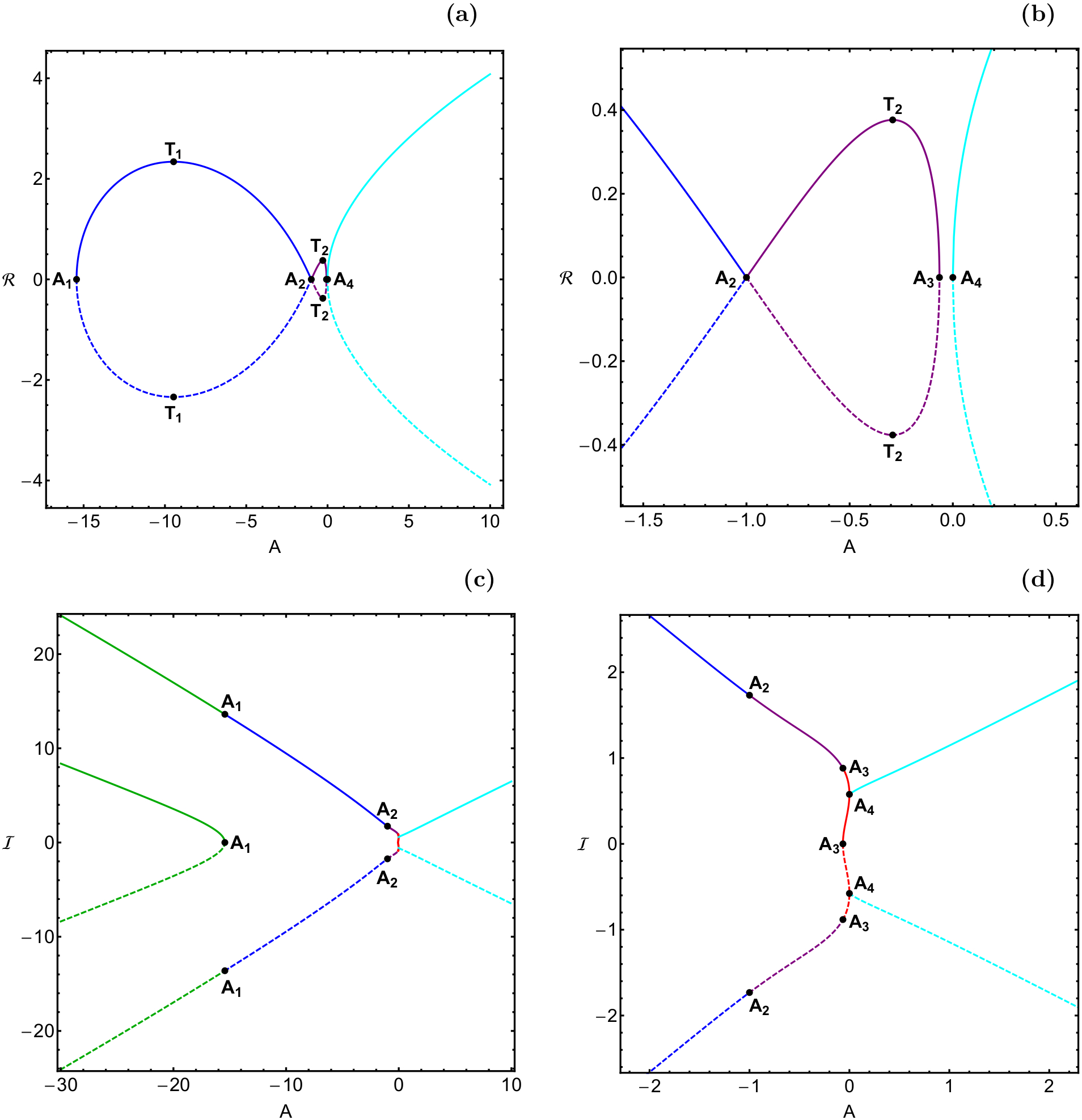}}
\caption{The space evolution of the real (a-upper left) and the imaginary (c-lower left) parts of the four roots $R_{ij}$, with $i = 1,...,5$, $j = 2,...,5$, when $A \in [-30,10]$. Panels (b) and (d) contain magnifications of the panels (a) and (c), respectively. When $A < A_1$ we have the roots $R_{12}$, $R_{13}$, $R_{14}$, and $R_{15}$ (green), when $A \in (A_1, A_2)$ we have the roots $R_{22}$, $R_{23}$, $R_{24}$, and $R_{25}$ (blue), when $A \in (A_2, A_3)$ we have the roots $R_{32}$, $R_{33}$, $R_{34}$, and $R_{35}$ (purple), when $A \in (A_3, A_4)$ we have the roots $R_{42}$, $R_{43}$, $R_{44}$, and $R_{45}$ (red), while when $A > A_4$ we have the roots $R_{52}$, $R_{53}$, $R_{54}$, and $R_{55}$ (cyan). The critical values of the oblateness coefficient ($A_1$, $A_2$, $A_3$, $A_4$) are pinpointed by black dots, while the points $T_1$ and $T_2$ indicate the turning points of the real parts. (Color figure online).}
\label{evol}
\end{figure*}

It would be very illuminating to determine how the positions of the four roots, on the complex plane, evolve as a function of the oblateness coefficient. Fig. \ref{evol} shows the parametric evolution of the positive real and imaginary parts of the four roots $R_{ij}$, $i = 1,...,5$, $j = 2,...,5$ on the complex plane, when $A \in [-40,10]$, with $\mathcal{R} = Re[\mathz]$ and $\mathcal{I} = Im[\mathz]$. When $A \to - \infty$ the imaginary roots $R_{12}$ and $R_{14}$ tend to collide to $+ \infty$, while the imaginary roots $R_{13}$ and $R_{15}$ tend to collide to $- \infty$. As we proceed to higher values of $A$ the imaginary roots $R_{12}$ and $R_{13}$ tend to the origin and for $A = A_1$ they collide and they mutually eliminated. For $A > A_1$ two real roots $R_{22}$ and $R{23}$ emerge from the origin and they start to move away from the center $(0,0)$. This behavior continuous up to $A = -9.46720$, while for higher values of the oblateness coefficient the tendency is reversed and the real roots start to come closer to the origin. When $A = A_2$ the two real roots collide at the origin and all three roots $R_{21}$, $R_{22}$, and $R_{23}$ completely disappear. For $A > A_2$ a new pair of real roots emerge from the origin and follow similar evolution, as the previous one. In particular, for $A_2 < A < -0.29149$ the two real roots move away from the origin, while for $-0.29149 < A < A_3$ they return back and when $A = A_3$ they collide, once more, with the central point $(0,0)$. As long as $A > A_3$ a new pair of pure imaginary roots emerge from the origin. As the value of the oblateness coefficient increases, thus tending to zero, the the four imaginary roots come closer as they move on collision course. Indeed, when $A = A_4$ $R_{42}$ and $R_{43}$ collide with $R_{44}$ and $R_{45}$, respectively thus annihilating each other, which implies that only the central root $\mathz = 0$ survives. Finally, when $A > A_4$ four complex conjugate roots emerge. It is interesting to note that the imaginary parts of the complex roots bifurcate exactly at the points where the four pure imaginary roots of the previous case collided. Our analysis indicates that with increasing value of the oblateness coefficient all four complex roots move away from the center, while their nature remains unperturbed when the primary bodies are oblate $(A > 0)$.

\section{The basins of convergence of the Newton-Raphson scheme}
\label{nrb}

The well-known Newton-Raphson optimal method of second provides one of the simplest ways for solving numerically an equation
with one variable. The corresponding iterative scheme read as
\begin{equation}
\mathz_{n+1} = \mathz_n - \frac{f(\mathz;A)_n}{f'(\mathz;A)_n} = \frac{9\mathz^3 \left(n_1 \mathz^4 + n_2 \mathz^2 + n_3 \right)}{d_1 \mathz^6 + d_2 \mathz^4 + d_3 \mathz^2 + d_4},
\label{nr}
\end{equation}
where $\mathz_n$ is the value of the $\mathz$ at the $n$-th step of the iterative process, while $f'(\mathz;A)$ is the first order derivative of $f(\mathz;A)$. Moreover, the analytical expressions of the numerical coefficients, entering the numerator and the denominator, are
\begin{align}
n_1 &= 18, \ \ \ n_2 = 6\left(2 + A \left(2A - 11 \right)\right), \nonumber\\
n_3 &= \left(2 + A \left(2A + 79\right)\right)a^2, \nonumber\\
d_1 &= 108, \ \ \ d_2 = 54\left(1 + \left(A - 10\right)A\right), \ \ \ d_3 = 648Aa, \nonumber\\
d_4 &= - \left(2 + A\left(2A + 31\right)\right)a^4.
\label{cfs}
\end{align}

The Newton-Raphson method works with the following philosophy: The code is activated with an initial complex number $\mathz = a + ib$, with $\mathcal{R} = a$ and $\mathcal{I} = b$, on the complex plane, while the iterative procedure continues until a root is reached, with the desired predefined accuracy. The numerical method converges for an initial condition $(\mathcal{R}, \mathcal{I})$, if the particular initial condition leads to one of the roots of the system. It is necessary to note that the Newton-Raphson method does not converge equally well for all the initial conditions on the complex plane. The Newton-Raphson basins of convergence or convergence areas/domains are composed of the sets of the initial conditions which lead to the same final state (root which acts as an numerical attractor). However, it should be clarified and emphasized that the Newton-Raphson basins of convergence should not be mistaken, by no means, with the basins of attractions which are present in dissipative system.

Looking the iterative formula of Eq. (\ref{nr}) we realize that the Newton-Raphson basins of convergence should reflect some of the most basic and intrinsic dynamical properties of the Hamiltonian system. This should be true because the iterative formula contains the equation of motion (\ref{eqmotz}) as well as its first order derivative.

To reveal the structures of the basins of convergence a double scan of the complex plan is performed. More precisely, a dense uniform grid of $1024 \times 1024$ $(\mathcal{R},\mathcal{I})$ nodes is defined, containing all the initial conditions which will be classified by the iterative scheme. The number $N$ of the iterations, required for obtaining the desired accuracy, is also monitored during the classification of the nodes. For our computations, the maximum allowed number of iterations is $N_{\rm max} = 500$. Moreover the iterations stop when a root is reached, with accuracy of $10^{-15}$ for both real and imaginary parts.

For the classification of the nodes on the complex plane we will use color-coded diagrams (CCDs), in which each pixel is assigned a different color, according to the final state (root) of the corresponding initial condition. Here we would like to clarify that the size of each CCD (or in other words the minimum and the maximum values of $\mathcal{R}$ and $\mathcal{I}$) is defined, in each case, in such a way so as to have a complete view of the overall geometry of the basins of convergence.

\begin{figure*}[!t]
\centering
\resizebox{0.80\hsize}{!}{\includegraphics{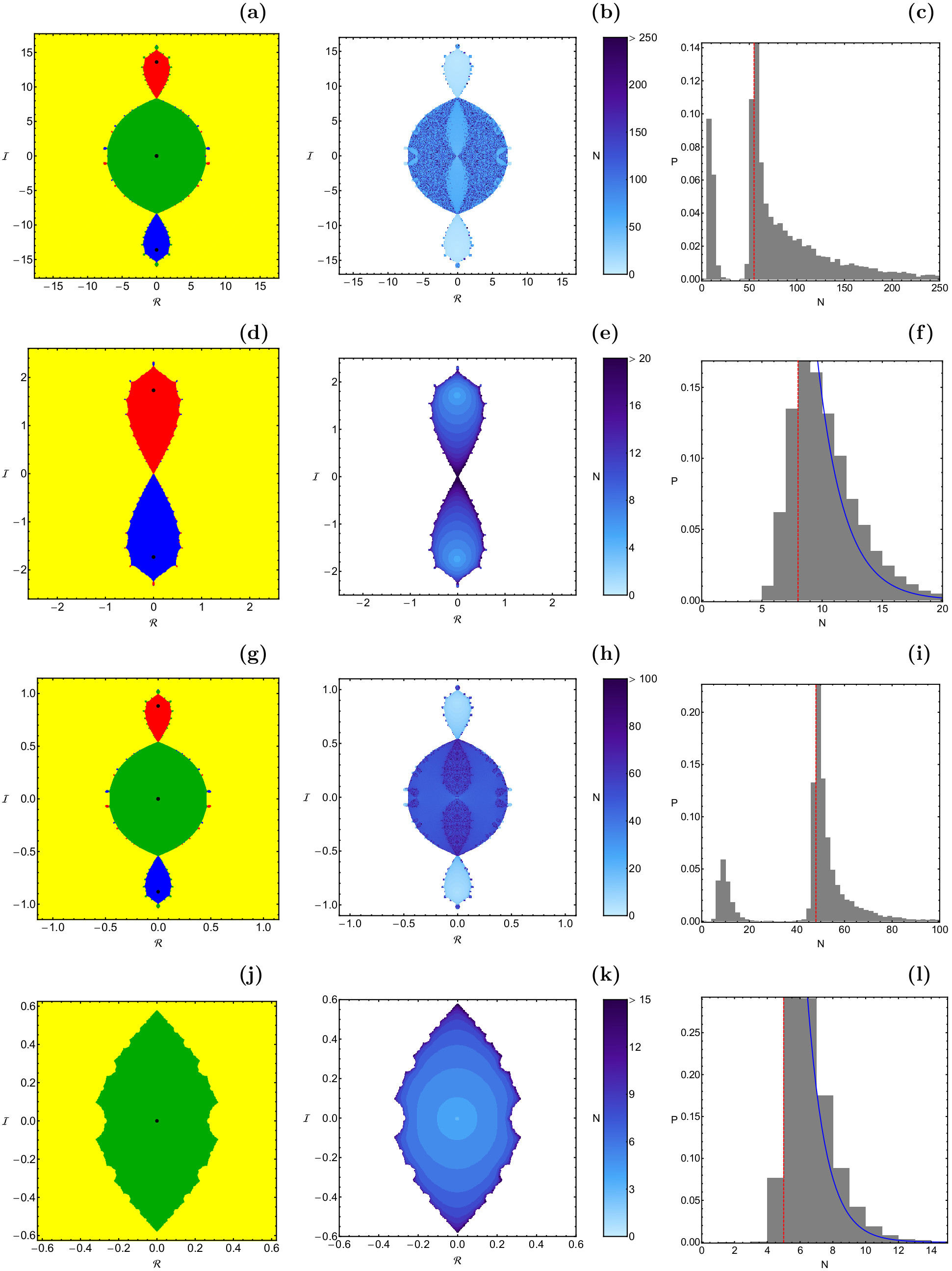}}
\caption{(First column): The Newton-Raphson basins of convergence on the complex plane for the critical values of the oblateness coefficient $A$. The color code, denoting the roots (black dots), is as follows: $R_1$ (green); $R_{2}$ (red); $R_{3}$ (blue); points tending to infinity (yellow); non-converging points (white). (Second column): The distribution of the corresponding number $N$ of required iterations. The non-converging points, as well those leading to infinity, are shown in white. (Third column): The corresponding probability distribution of required iterations. The vertical, dashed, red line indicates, in each case, the most probable number $N^{*}$ of iterations. (First row): $A = A_1$; (Second row): $A = A_2$; (Third row): $A = A_3$; (Fourth row): $A = A_4$. (Color figure online).}
\label{c0}
\end{figure*}

The Newton-Raphson basins of convergence, corresponding to the four critical values of the oblateness coefficient, are presented in the first column of Fig. \ref{c0}. It is evident that in all cases the area of all the types of the basins of convergence is finite. Moreover, when $A = A_1$ or $A = A_2$ the convergence region of the central root $\mathz = 0$ has an elliptic shape. On the other hand, the basins of convergence associated with the two pure imaginary roots form mainly two lobes, attached to the central region. When $A = 0$ only one unified basin of convergence is present which has a rhomboidal shape. Furthermore, it is seen that the vast majority of the complex plane is covered by initial conditions which do not converge to any of the roots (yellow regions). Additional numerical calculations indicate that for all these initial conditions the Newton-Raphson iterative scheme leads progressively to extremely large real or imaginary numbers. This behavior is in fact a numerical indication that for these initial conditions the Newton-Raphson iterator leads asymptotically to infinity.

We see that the regions in the vicinity of the basin boundaries are highly fractal\footnote{By the term fractal we simply mean that the particular area has a fractal-like geometry, without conducting, at least for now, any additional calculations for computing the degree of fractality, as in \cite{AVS01,AVS09}.}, which implies that the final state (root) of an initial condition inside this area is highly sensitive. More precisely, even the slightest change of the initial conditions automatically leads to a completely different root, which is a classical indication of chaos. Therefore, for the initial conditions in the basin boundaries it is almost impossible to predict their final states (roots).

In the second column of the same figure the distribution of the corresponding number $(N)$ of iterations required for obtaining the desired accuracy is given, using tones of blue. We observe that this type of diagrams unveils hidden patterns, regarding the geometry of the convergence regions. One may observe that when $A = A_1$ or $A = A_3$ inside the elliptic region, corresponding to the central root $\mathz = 0$, there are two more lobes, indicating the additional two pure imaginary roots that have been mutually annihilated at the origin, thus merging with the central root. For $A = A_1$ we seen in panel (b) that the distribution of iterations of the two inner lobes is very smooth, while on the other hand the distribution of iterations in the remaining of the central basin is very noisy. In panel (h), where $A = A_3$, we observe the exact opposite phenomenon, that is noisy distribution of iterations inside the lobes and smooth distribution of iterations inside the remaining central region. Furthermore, for both cases ($A = A_1$ and $A = A_3$), the Newton-Raphson iterative scheme requires, in average, much more iterations for the initial conditions which lead to $\mathz = 0$, in relation to the required iterations for the initial conditions which lead to one of the pure imaginary roots.

The corresponding probability distribution of the required iterations is given in the third column of Fig. \ref{c0}. The definition of the probability $P$ is the following: if $N_0$ complex initial conditions $(\mathcal{R},\mathcal{I})$ converge, after $N$ iterations, to one of the roots then $P = N_0/N_t$, where $N_t$ is the total number of nodes in every CCD. In all plots the tails of the histograms extend so as to cover 98\% of the corresponding distributions of iterations. The vertical, red, dashed line in the probability histograms denote the most probable number $N^{*}$ of iterations, while the blue lines in the histograms indicate the best fit (if possible) to the right-hand side $N > N^{*}$ of them (more details regarding the best fit are given in the following subsection \ref{geno}). Panels (c) and (i) reveal that when $A = A_1$ or $A = A_3$ the histograms are composed of two disjoint parts. In fact with this behavior the histograms confirm the phenomenon observed earlier in the diagrams showing the distribution of iterations. More precisely, the first small part of these two histograms corresponds to the initial conditions which form the two lobes and converge relatively fast. On the contrary, the second and main body of the histograms corresponds to the initial conditions which lead to the central root, for which the Newton-Raphson iteration requires a substantial amount of iterations for obtaining the desired accuracy.

Our computations suggest that when $A = -1$ (see the second column of Fig. \ref{c0}) and $A = 0$ (see the fourth column of Fig. \ref{c0}) both distributions of iterations and probability display the normal and expected behavior. So far, we do not have a definitive answer explaining the strange and unexpected behavior observed for $A = A_1$ and $A = A_3$. However, we assume that a partial answer to this strange behavior should be the fact that these two values of the oblateness coefficient are critical values, at which equilibrium points (roots) are mutually annihilated.

In the following subsections we will determine how the oblateness coefficient $A$ affects the structure of the Newton-raphson basins of convergence in the Sitnikov four-body problem, by considering several cases, regarding the nature of the five roots.

\begin{figure*}[!t]
\centering
\resizebox{\hsize}{!}{\includegraphics{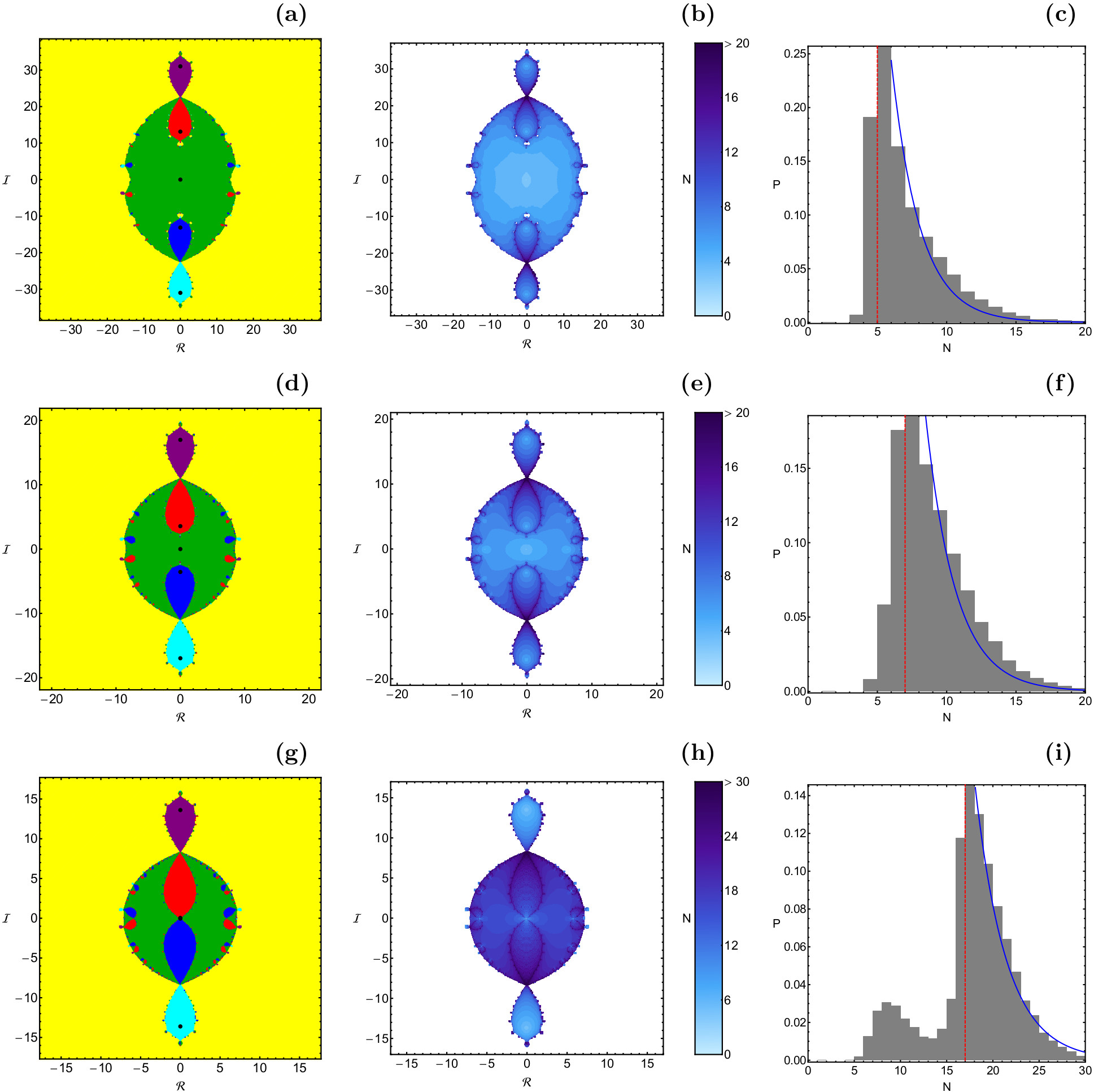}}
\caption{(First column): The Newton-Raphson basins of convergence on the complex plane for the first case, where $A < A_1$. The color code, denoting the roots (black dots), is as follows: $R_{11}$ (green); $R_{12}$ (red); $R_{13}$ (blue); $R_{14}$ (purple); $R_{15}$ (cyan); points tending to infinity (yellow); non-converging points (white). (Second column): The distribution of the corresponding number $N$ of required iterations. The non-converging points, as well those leading to infinity, are shown in white. (Third column): The corresponding probability distribution of required iterations. The vertical, dashed, red line indicates, in each case, the most probable number $N^{*}$ of iterations. (First row): $A = -40$; (Second row): $A = -20$; (Third row): $A = -15.436$. (Color figure online).}
\label{c1}
\end{figure*}

\begin{figure*}[!t]
\centering
\resizebox{\hsize}{!}{\includegraphics{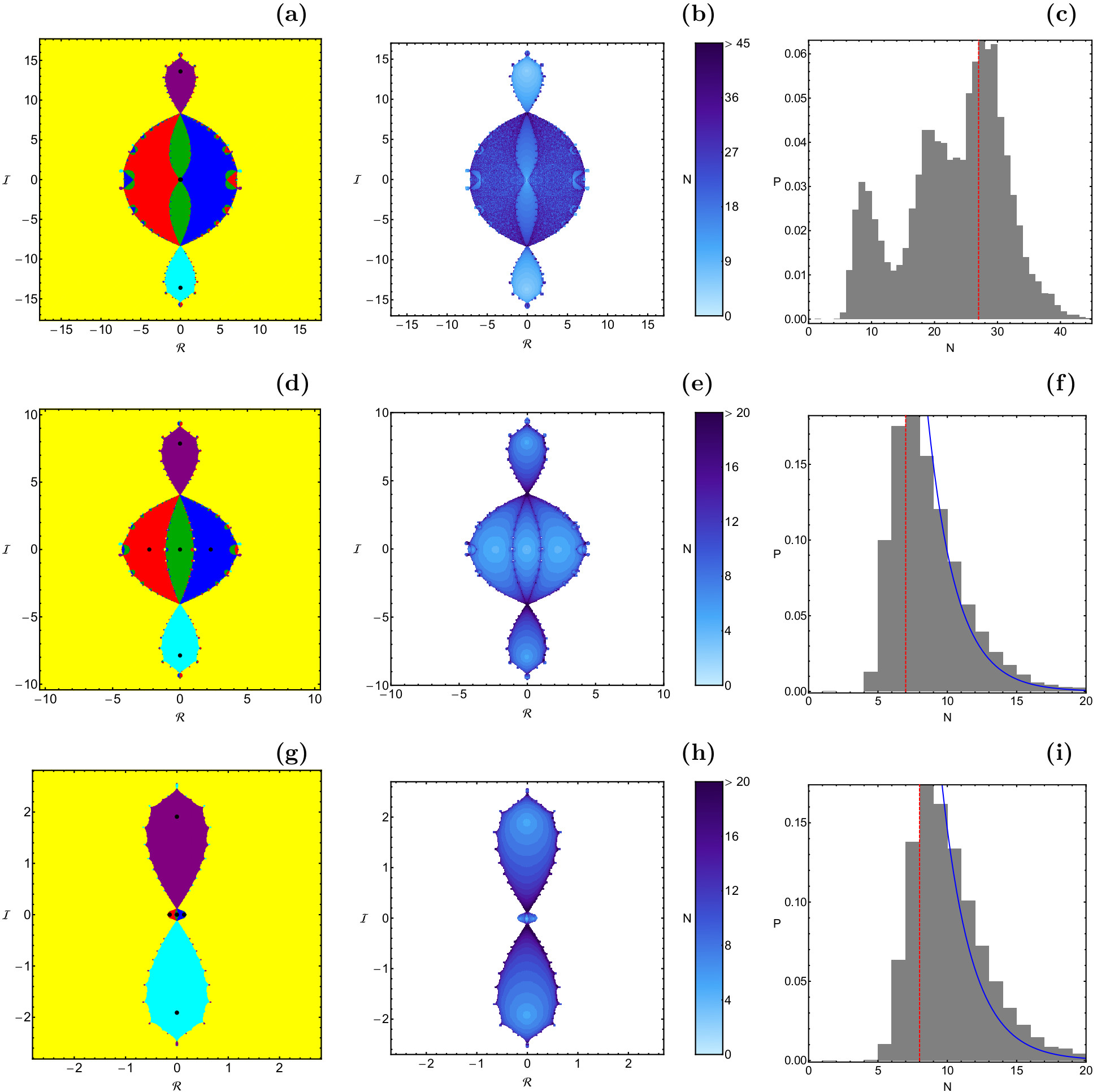}}
\caption{(First column): The Newton-Raphson basins of convergence on the complex plane for the second case, where $A_1 < A < A_2$. The color code, denoting the roots (black dots), is as follows: $R_{21}$ (green); $R_{22}$ (red); $R_{23}$ (blue); $R_{24}$ (purple); $R_{25}$ (cyan); points tending to infinity (yellow); non-converging points (white). (Second column): The distribution of the corresponding number $N$ of required iterations. The non-converging points, as well those leading to infinity, are shown in white. (Third column): The corresponding probability distribution of required iterations. The vertical, dashed, red line indicates, in each case, the most probable number $N^{*}$ of iterations. (First row): $A = -15.434$; (Second row): $A = -8$; (Third row): $A = -1.2$. (Color figure online).}
\label{c2}
\end{figure*}

\begin{figure*}[!t]
\centering
\resizebox{\hsize}{!}{\includegraphics{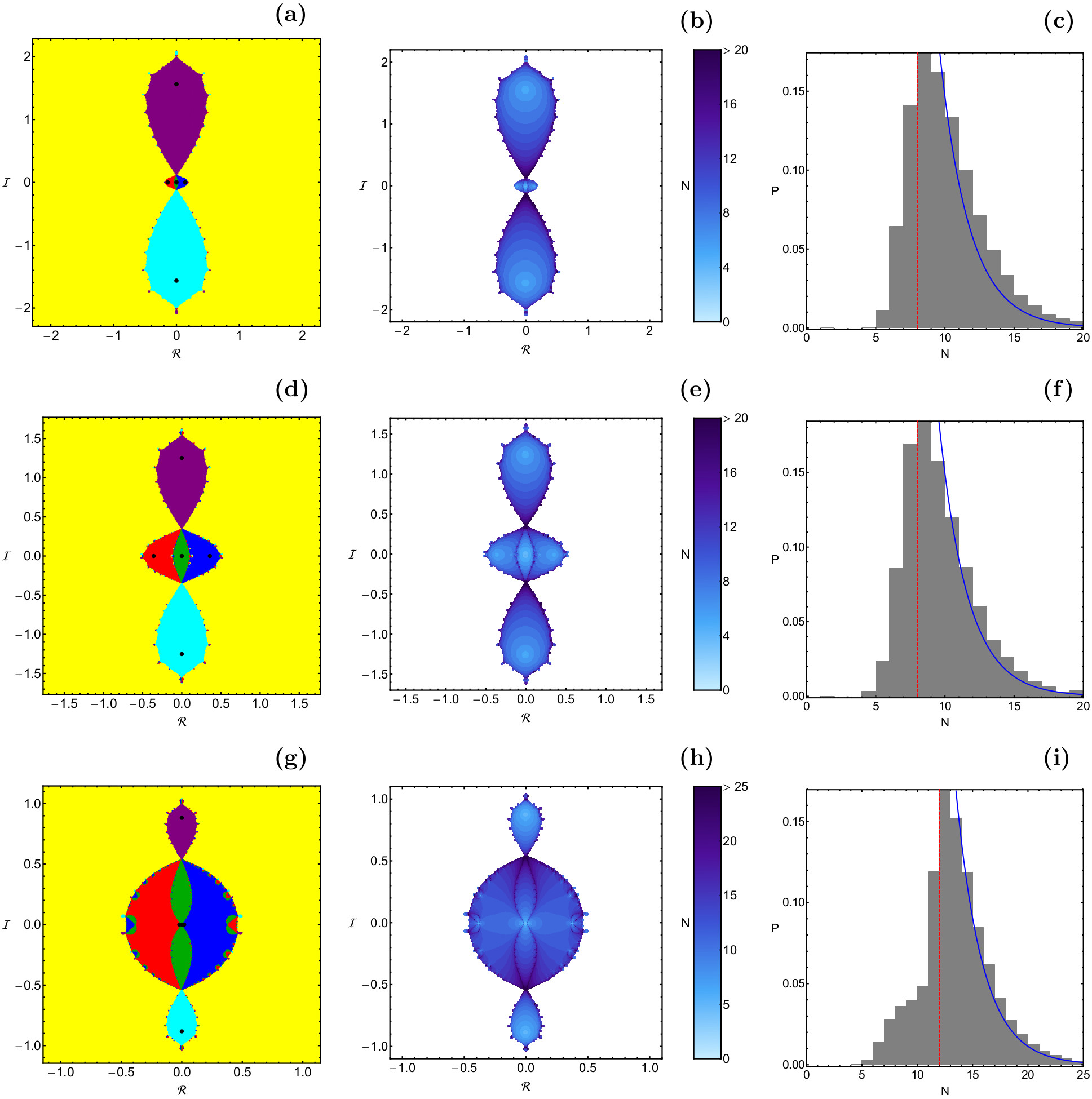}}
\caption{(First column): The Newton-Raphson basins of convergence on the complex plane for the third case, where $A_2 < A < A_3$. The color code, denoting the roots (black dots), is as follows: $R_{31}$ (green); $R_{32}$ (red); $R_{33}$ (blue); $R_{34}$ (purple); $R_{35}$ (cyan); points tending to infinity (yellow); non-converging points (white). (Second column): The distribution of the corresponding number $N$ of required iterations. The non-converging points, as well those leading to infinity, are shown in white. (Third column): The corresponding probability distribution of required iterations. The vertical, dashed, red line indicates, in each case, the most probable number $N^{*}$ of iterations. (First row): $A = -0.8$; (Second row): $A = -0.4$; (Third row): $A = -0.065$. (Color figure online).}
\label{c3}
\end{figure*}

\begin{figure*}[!t]
\centering
\resizebox{\hsize}{!}{\includegraphics{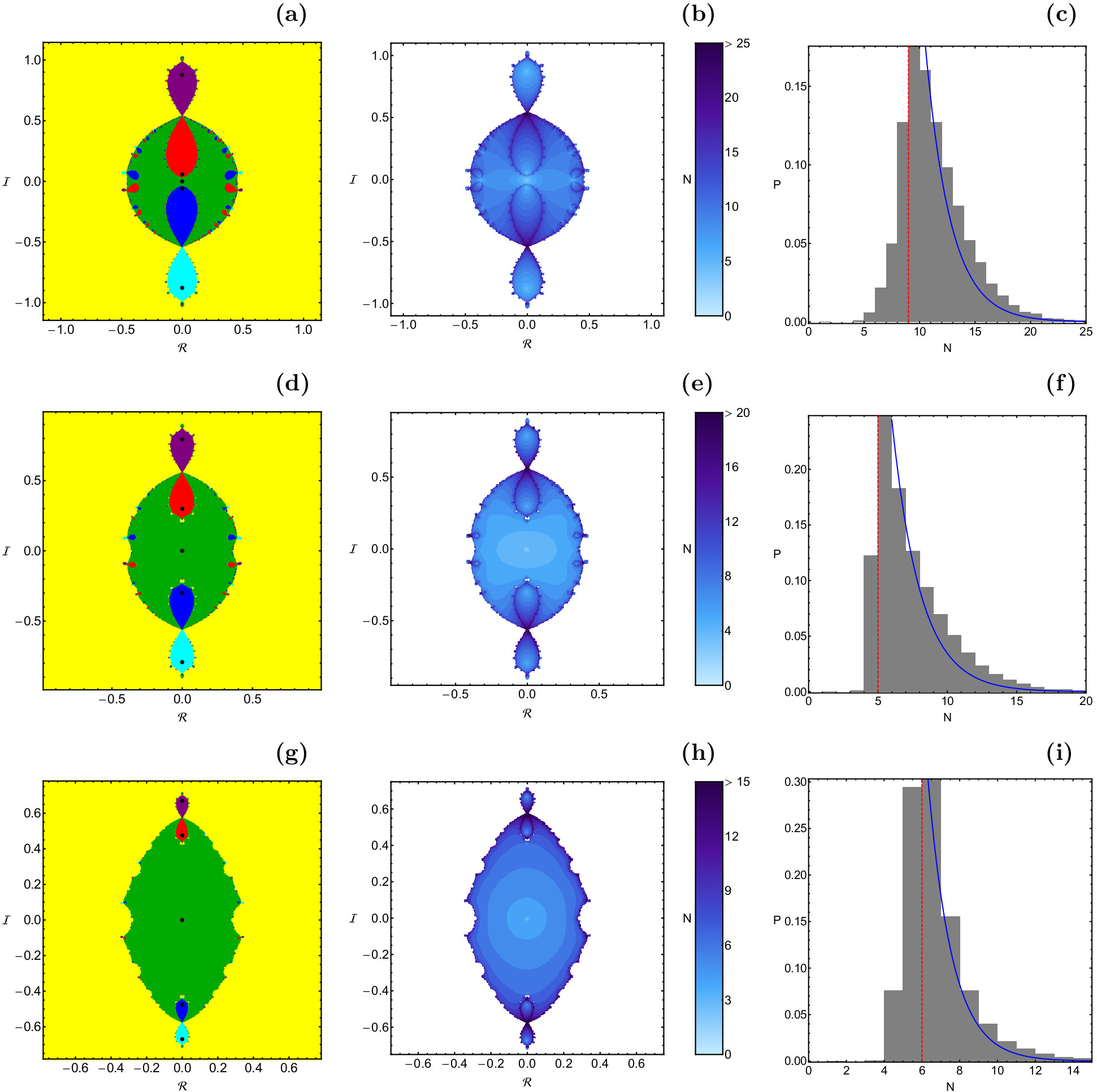}}
\caption{(First column): The Newton-Raphson basins of convergence on the complex plane for the fourth case, where $A_3 < A < A_4$. The color code, denoting the roots (black dots), is as follows: $R_{41}$ (green); $R_{42}$ (red); $R_{43}$ (blue); $R_{44}$ (purple); $R_{45}$ (cyan); points tending to infinity (yellow); non-converging points (white). (Second column): The distribution of the corresponding number $N$ of required iterations. The non-converging points, as well those leading to infinity, are shown in white. (Third column): The corresponding probability distribution of required iterations. The vertical, dashed, red line indicates, in each case, the most probable number $N^{*}$ of iterations. (First row): $A = -0.063$; (Second row): $A = -0.03$; (Third row): $A = -0.005$. (Color figure online).}
\label{c4}
\end{figure*}

\begin{figure*}[!t]
\centering
\resizebox{\hsize}{!}{\includegraphics{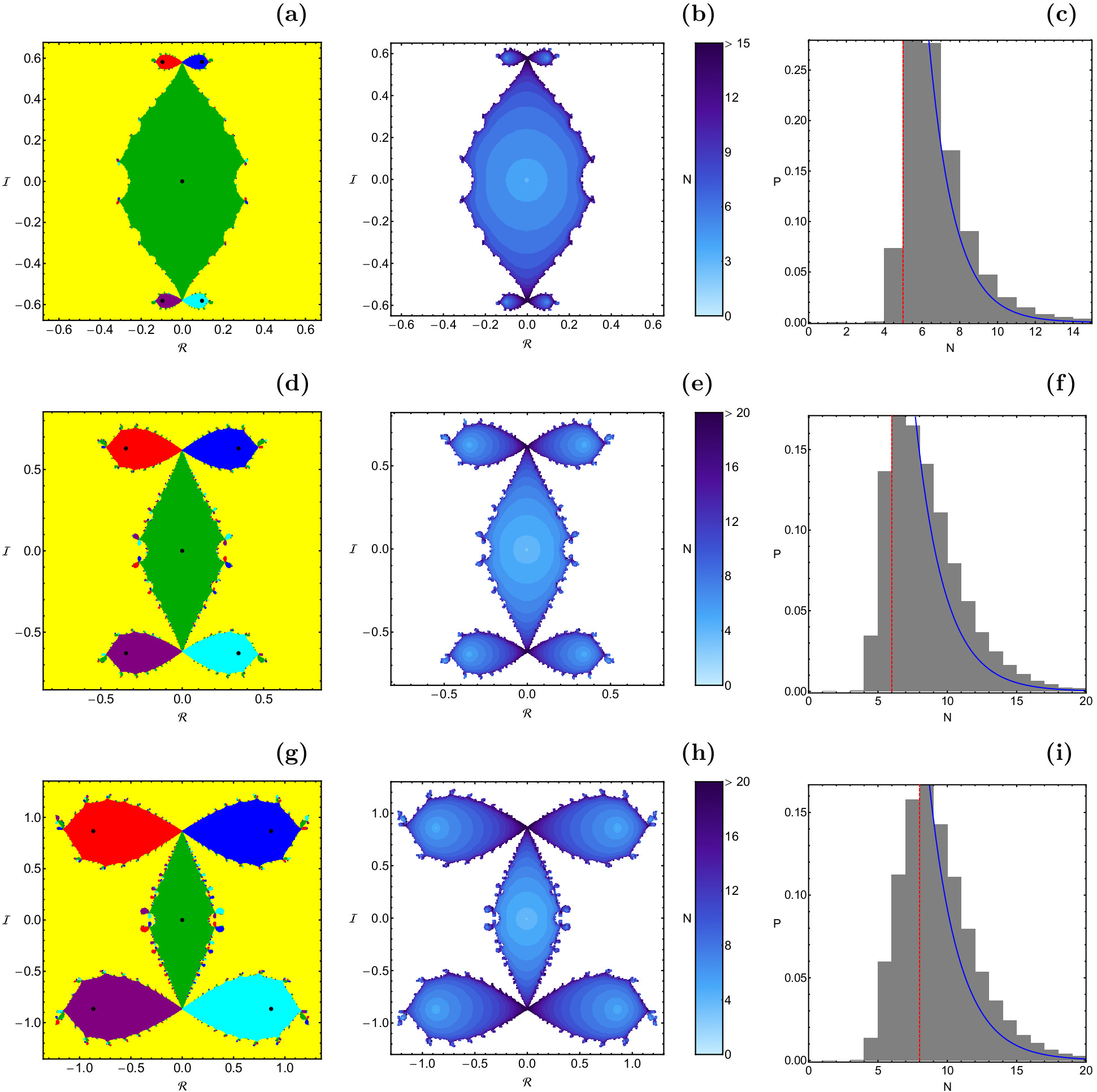}}
\caption{(First column): The Newton-Raphson basins of convergence on the complex plane for the fifth case, where $A > A_4$. The color code, denoting the roots (black dots), is as follows: $R_{51}$ (green); $R_{52}$ (red); $R_{53}$ (blue); $R_{54}$ (purple); $R_{55}$ (cyan); points tending to infinity (yellow); non-converging points (white). (Second column): The distribution of the corresponding number $N$ of required iterations. The non-converging points, as well those leading to infinity, are shown in white. (Third column): The corresponding probability distribution of required iterations. The vertical, dashed, red line indicates, in each case, the most probable number $N^{*}$ of iterations. (First row): $A = 0.005$; (Second row): $A = 0.07$; (Third row): $A = 0.5$. (Color figure online).}
\label{c5}
\end{figure*}

\subsection{Case I: $A < A_1$}
\label{ss1}

We begin with the first case, where the equation $f(\mathz;A) = 0$ has, apart from the $\mathz = 0$ root, four pure imaginary roots. The Newton-Raphson basins of convergence on the complex plane, for three values of the oblateness coefficient, are illustrated in the first column of Fig. \ref{c1}. In the second column of the same figure we present the corresponding number $N$ of iterations, while the corresponding probability distribution of the required iterations is given in the third column of Fig. \ref{c1}.

With increasing value of the oblateness coefficient the most important phenomena which take place are the following:
\begin{enumerate}
  \item The area of the basins of convergence, with a lobe shape, which correspond to the four pure imaginary roots increases, while at the same time the two inner lobes, corresponding to roots $R_{12}$ and $R_{13}$, come closer.
  \item The shape of the central convergence region, corresponding to root $R_{11} = \mathz = 0$, changes from elliptical to almost circular.
  \item The extent of the secondary basins of convergence that are present in the vicinity of the boundaries of the main basins increases.
\end{enumerate}

In panel (h) of Fig. \ref{c1} we observe that when $A = -15.436$, that is a value of the oblateness coefficient just before the first critical value, the required iterations for reaching the roots $R_{11}$, $R_{12}$ and $R_{13}$ are substantially higher that those required for reaching the imaginary roots $R_{14}$ and $R_{15}$. Indeed, in panel (i) of the same figure we see that the histogram is composed of two parts corresponding to fast and slow converging points. In fact, this situation is very similar to that observed earlier in Fig. \ref{c0}, regarding the critical values. We suspect that the presence of such slow converging points for $A = -15.436$ is directly related with the fact that at $A = A_1$ the two imaginary roots $R_{12}$ and $R_{13}$ collide at the center and they are mutually eliminated.

\subsection{Case II: $A_1 < A < A_2$}
\label{ss2}

The next case under consideration involves the scenario where there are two real and two pure imaginary roots, along with the $\mathz = 0$ root. In the first column of Fig. \ref{c2} we present the Newton-Raphson basins of convergence for three values of the oblateness coefficient. The second and third column of Fig. \ref{c2} contain the corresponding number $N$ of iterations, and the probability distribution of the required iterations, respectively.

As we proceed to higher values of $A$ the main changes, regarding the geometry of the convergence areas, are the following:
\begin{enumerate}
  \item Initially the basins of convergence, corresponding to the central root $z = 0$, are composed of two elongated lobes, which later on they merge thus forming a unified rhomboidal area.
  \item The area of the lobed basins of convergence, corresponding to the imaginary roots $R_{24}$ and $R_{25}$ expands rapidly.
  \item The extent of the convergence regions, corresponding to roots $R_{21}$, $R_{22}$ and $R_{23}$ decreases, very quickly.
\end{enumerate}

In panel (b) of Fig. \ref{c2} we see that the distribution of iterations, corresponding to the two real roots $R_{22}$ and $R_{23}$ is very noisy. Furthermore, according to panel (c) of the same figure, the average number of required iterations for the two real roots is about three times higher that the number of required iterations for the rest of the roots. Again, this strange behavior should be related with the fact that $A = -15.434$ is just above the critical value $A_1$, where the dynamical properties of the system change drastically.

\subsection{Case III: $A_2 < A < A_3$}
\label{ss3}

We continue with the third case, where there are, once more, two real and two pure imaginary roots, along with the universal $(0,0)$ root. The convergence regions on the complex plane, for three values of the oblateness coefficient, are illustrated in the first column of Fig. \ref{c3}, while the corresponding distributions of iterations and probability are presented in the second and third column of the same figure, respectively.

It is evident that the evolution of the geometry of the basins of convergence follows the exact opposite path of the previous studied case.

\subsection{Case IV: $A_3 < A < A_4$}
\label{ss4}

In the first column of Fig. \ref{c4} we depict the Newton-Raphson basins of convergence for three values of the oblateness coefficient, when four pure imaginary roots are present, along with the $\mathz = 0$ root. The corresponding distributions of the required iterations and the probability are given in the second and third column of Fig. \ref{c4}, respectively.

As we proceed to higher values of the oblateness coefficient the geometry of the convergence domains changes as follows:
\begin{enumerate}
  \item The two pairs of lobed basins of convergence, corresponding to the four pure imaginary roots $R_{42}$, $R_{43}$, $R_{44}$, and $R_{45}$ move away from each other, while at the same time their area is heavily reduced.
  \item The extent of the central convergence region, associated with the root $\mathz = 0$ increases, while its shape changes from circular to rhomboidal.
  \item All the secondary basins, that initially are present at the boundaries of the central region, disappear, thus reducing the degree of fractality of the complex plane.
\end{enumerate}

Looking carefully at Fig. \ref{c4} we may argue that the evolution of the geometry of the basins of convergence in this case is, in general terms, opposite with respect to that we seem earlier in Fig. \ref{c1}, when $A < A_1$.

\subsection{Case V: $A > A_4$}
\label{ss5}

Our exploration ends with the case where there are four complex conjugate roots, along with the $\mathz = 0$ root. The Newton-Raphson basins of convergence, for three values of the oblateness coefficient $A$, are presented in the first column of Fig. \ref{c5}. The corresponding number $N$ of iterations, and the probability distribution of the required iterations are give in the second and third column of Fig. \ref{c5}, respectively.

During the transition from the case where $A_3 < A < A_4$ to the case where $A > A_4$ it is evident that the orientation of the four lobed basins of convergence changes from vertical (which implies the presence of four pure imaginary roots) to horizontal (which suggests the existence of four complex roots).

As the primary bodies become more oblate the most important changes that occur, regarding the geometry of the basins of convergence are:
\begin{enumerate}
  \item The area of the lobed convergence regions, corresponding to complex roots $R_{52}$, $R_{53}$, $R_{54}$, and $R_{55}$ increases rapidly.
  \item The extent of the central convergence area, corresponding to the root $\mathz = 0$ decreases.
  \item All the secondary basins of convergence, which are located at the boundaries of the main basins, become more prominent, which suggests that the degree of fractality of the complex plane increases.
\end{enumerate}

\subsection{An overview analysis}
\label{geno}

\begin{figure*}[!t]
\centering
\resizebox{\hsize}{!}{\includegraphics{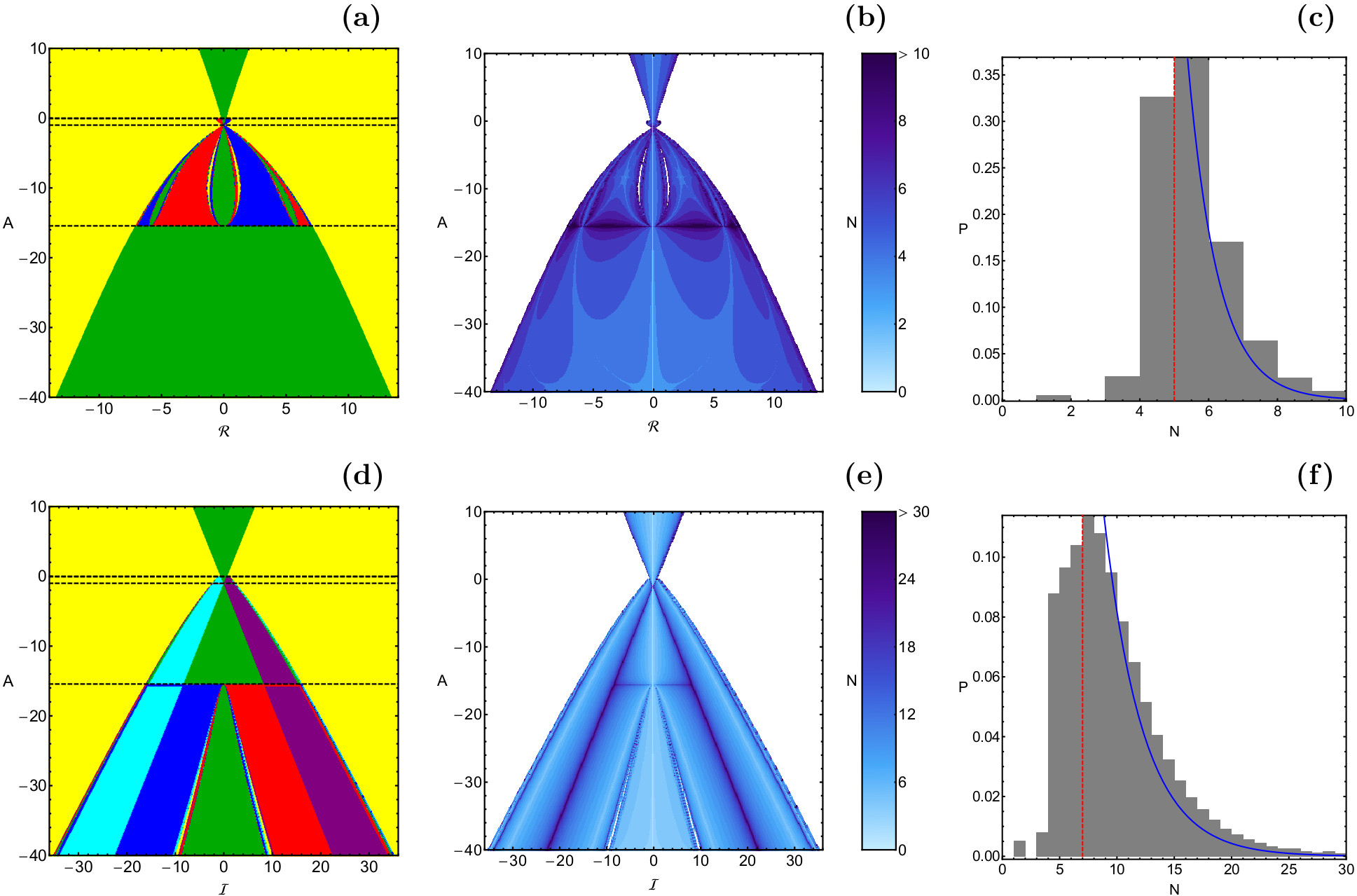}}
\caption{(First row): The $(\mathcal{R}, A)$ plane. (Second row): The $(\mathcal{I}, A)$ plane. (First column): The Newton-Raphson basins of convergence, when $A \in [-40,10]$. The color code, denoting the five roots, is as in the previous cases. (Second column): The distribution of the corresponding number $N$ of required iterations. The non-converging points, as well those leading to infinity, are shown in white. (Third column): The corresponding probability distribution of required iterations. The horizontal, dashed, black lines indicate the critical values of the oblateness coefficient. (Color figure online)}.
\label{ria}
\end{figure*}

The color-coded diagrams (CCDs) on the complex plane, presented earlier in subsections \ref{ss1}, \ref{ss2}, \ref{ss3}, \ref{ss4}, and \ref{ss5} provide sufficient information regarding the convergence domains, however for only a fixed value of the oblateness coefficient $A$. In order to overcome this drawback we can define a new type of distribution of initial conditions which will allow us to scan a continuous spectrum of $A$ values, rather than few discrete levels. The most interesting configuration is to set either the real part or the imaginary part equal to zero, while the value of the oblateness coefficient will vary in the interval $[-40,10]$. This technique allows us to construct, once more, a two-dimensional plane in which the $\mathcal{R}$ or the $\mathcal{I}$ is the abscissa, while the value of $A$ is always the ordinate. Panels (a) and (d) of Fig. \ref{ria} show the basins of convergence on the $(\mathcal{R},A)$ and $(\mathcal{I},A)$ planes, respectively. The corresponding distributions of iterations and probability are given in panels (b), (c) and (e), (f), respectively. It is interesting to observe in panel (b) the structures created inside the unified green region, corresponding to the central root $(\mathz = 0)$. Evidently, the geometry of the convergence structure changes exactly at the critical values.

Additional interesting information could be extracted from the probability distributions of iterations presented in the third row of Fig. \ref{ria}. In particular, it would be very interesting to try to obtain the best fit of the tails\footnote{By the term ``tails" of the distributions we refer to the right-hand side of the histograms, that is, for $N > N^{*}$.} of the distributions. For fitting the tails of the histograms, we used the Laplace distribution, which is the most natural choice, since this type of distribution is very common in systems displaying transient chaos (see e.g., \cite{ML01,SASL06,SS08}). Our calculations strongly indicate that in the vast majority of the cases the Laplace distribution is the best fit to our data. The only cases where the Laplace distribution fails to properly fit the corresponding numerical data is the cases corresponding to $A = \{A_1, -15.434, A_2\}$, where the corresponding histograms display several peaks.

The probability density function (PDF) of the Laplace distribution is given by
\begin{equation}
P(N | a,b) = \frac{1}{2b}
 \begin{cases}
      \exp\left(- \frac{a - N}{b} \right), & \text{if } N < a \\
      \exp\left(- \frac{N - a}{b} \right), & \text{if } N \geq a
 \end{cases},
\label{pdf}
\end{equation}
where $a$ is the location parameter, while $b > 0$, is the diversity. In our case we are interested only for the $x \geq a$ part of the distribution function.

In Table \ref{t1} we present the values of the location parameter $a$ and the diversity $b$, as they have been obtained through the best fit, for all cases discussed in the previous subsections. One may observe that for most of the cases the location parameter $a$ is very close to the most probable number $N^{*}$ of iterations, while in some cases these two quantities coincide.

\begin{table}[!t]
\begin{center}
   \caption{The values of the location parameter $a$ and the diversity $b$, related to the most probable number $N^{*}$ of iterations, for all the studied cases shown earlier in the CCDs.}
   \label{t1}
   \setlength{\tabcolsep}{10pt}
   \begin{tabular}{@{}lrrrr}
      \hline
      Figure & $A$ & $N^{*}$ & $a$ & $b$ \\
      \hline
      \ref{c0}c  &   $A_1$ & 55 &           - &    - \\
      \ref{c0}f  &   $A_2$ &  8 & $N^{*} + 1$ & 2.23 \\
      \ref{c0}i  &   $A_3$ & 48 & $N^{*} + 1$ &    - \\
      \ref{c0}l  &   $A_4$ &  5 & $N^{*} + 1$ & 1.15 \\
      \hline
      \ref{c1}c  &    -40  &  5 & $N^{*} + 1$ & 2.05 \\
      \ref{c1}f  &    -20  &  7 & $N^{*} + 1$ & 2.16 \\
      \ref{c1}i  & -15.436 & 17 & $N^{*} + 1$ & 3.36 \\
      \hline
      \ref{c2}c  & -15.434 & 27 &           - &    - \\
      \ref{c2}f  &      -8 &  7 & $N^{*} + 1$ & 2.05 \\
      \ref{c2}i  &    -1.2 &  8 & $N^{*} + 1$ & 2.15 \\
      \hline
      \ref{c3}c  &    -0.8 &  8 & $N^{*} + 1$ & 2.13 \\
      \ref{c3}f  &    -0.4 &  8 & $N^{*} + 1$ & 2.06 \\
      \ref{c3}i  &  -0.065 & 12 & $N^{*} + 1$ & 2.40 \\
      \hline
      \ref{c4}c  &  -0.063 &  9 & $N^{*} + 1$ & 2.25 \\
      \ref{c4}f  &   -0.03 &  5 & $N^{*} + 1$ & 2.04 \\
      \ref{c4}i  &  -0.005 &  6 & $N^{*}$     & 1.29 \\
      \hline
      \ref{c5}c  &   0.005 &  5 & $N^{*} + 1$ & 1.37 \\
      \ref{c5}f  &    0.07 &  6 & $N^{*} + 1$ & 2.11 \\
      \ref{c5}i  &     0.5 &  8 & $N^{*}$     & 2.15 \\
      \hline
      \ref{ria}c &       - &  5 & $N^{*}    $ & 0.87 \\
      \ref{ria}f &       - &  7 & $N^{*} + 1$ & 3.39 \\
      \hline
   \end{tabular}
\end{center}
\end{table}

\section{Parametric evolution of the basin entropy}
\label{pebe}

So far, in the numerical results presented in the previous Section, we used only qualitative arguments for discussing the degree of the fractality of the basins of convergence on the complex plane. There is no doubt that quantitative results, regarding the evolution of the fractality, would be very informative. Very recently, in \cite{DWGGS16}, a new quantitative tool was introduced, for measuring the degree of the basin fractality. This new dynamical quantity is called ``basin entropy" and it measures the degree of fractality (or unpredictability) of the basins, by examining their topological properties.

The basin entropy works according to the following numerical algorithm. If there are $N(A)$ attractors (equilibrium points or roots) in a certain region $R$ on the complex plane, then we subdivide $R$ into a grid of $N$ square boxes, where each cell of the gird may contain between 1 and $N(A)$ attractors. Then the probability that inside the cell $i$ the corresponding attractor is $j$ is denoted by $P_{i,j}$. Taking into account that inside each cell the initial conditions are completely independent, the Gibbs entropy, of every cell $i$ reads
\begin{equation}
S_{i} = \sum_{j=1}^{m_{i}}P_{i,j}\log_{10}\left(\frac{1}{P_{i,j}}\right),
\end{equation}
where $m_{i} \in [1,N_{A}]$ is the total number of the attractors inside the cell $i$.

The total entropy of the entire region $R$, on the complex plane, can easily be calculated by adding the entropies of the $N$ cells of the grid as $S = \sum_{i=1}^{N} S_{i}$. Therefore, the total entropy, corresponding to the total number of cells $N$ is called basin entropy and it is given by
\begin{equation}
S_{b} = \frac{1}{N}\sum_{i=1}^{N}\sum_{j=1}^{m_{i}}P_{i,j}\log_{10}\left(\frac{1}{P_{i,j}}\right).
\end{equation}

\begin{figure*}[!t]
\centering
\resizebox{\hsize}{!}{\includegraphics{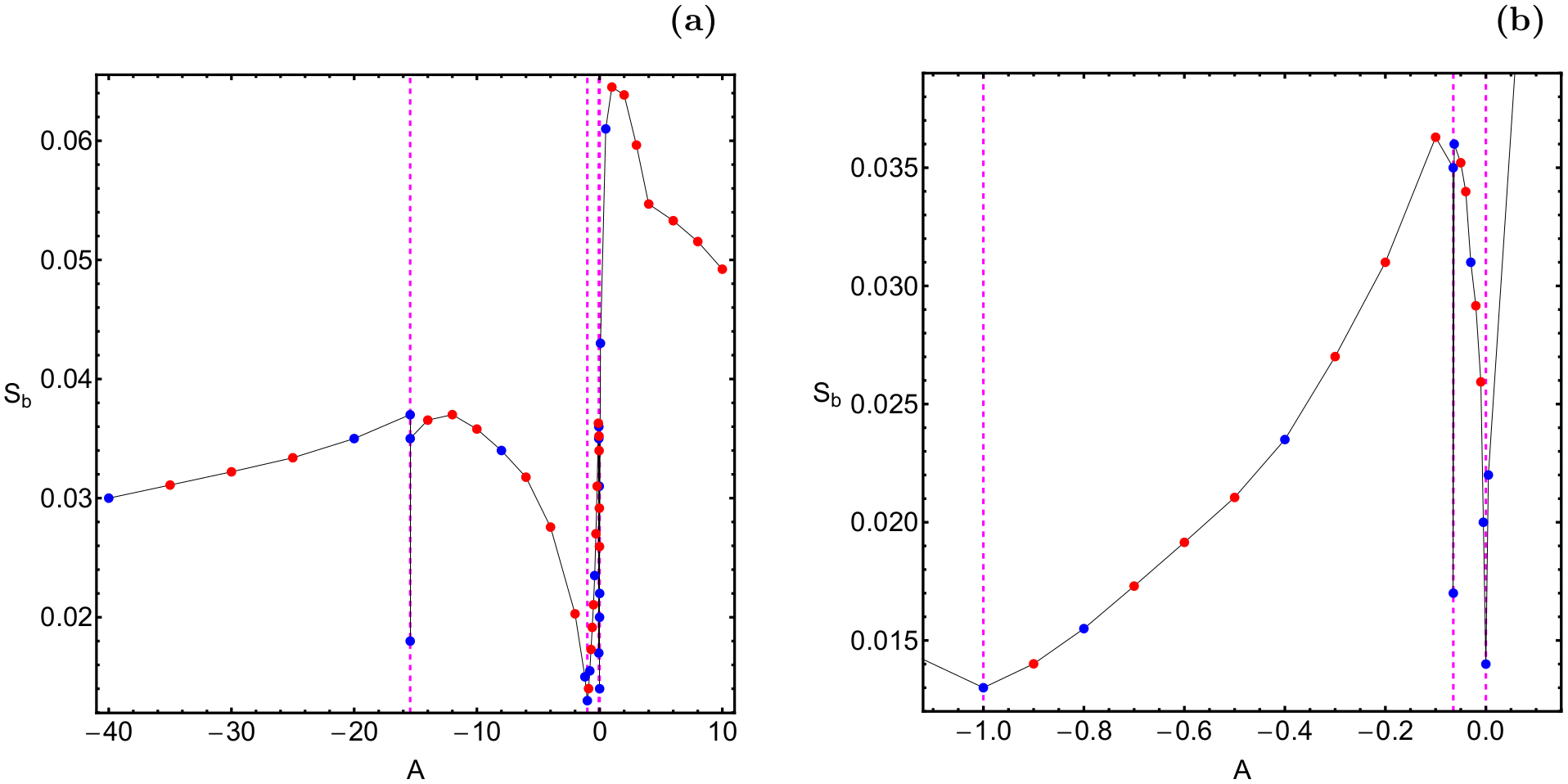}}
\caption{(a-left): Evolution of the basin entropy $S_b$, of the complex plane, as a function of the oblateness coefficient $A$. (b-right): Magnification of panel (a). The vertical, dashed, magenta lines indicate the critical values of $A$. Blue dots correspond to cases (values of A), for which the corresponding CCD is presented in Section \ref{nrb}, while red dots suggest cases, whose CCDs have not been displayed in the same Section. (Color figure online).}
\label{be}
\end{figure*}

Following the above-mentioned algorithm and also using the value $\varepsilon = 0.005$, suggested in \cite{DWGGS16}, we calculated the numerical value of the basin entropy $S_b$ of the complex plane, for several values of the oblateness coefficient $A$. At this point, it should be emphasized that the initial conditions, on the complex plane, for which the Newton-Raphson iterative scheme leads to infinity were counted as an additional type of basin, which coexist along with the regular basins of convergence, associated with the roots of the system. Fig. \ref{be}(a-b) illustrates the parametric evolution of the basins entropy, as a function of the oblateness coefficient. Here it should be noted that for this diagram we used results not only from the cases, of Figs. \ref{c0}, \ref{c1}, \ref{c2}, \ref{c3}, \ref{c4}, and \ref{c5}, but also from additional levels of the oblateness coefficient.

Looking at both panels of Fig. \ref{be} it becomes evident that:
\begin{itemize}
  \item When $A \to A_1$ the basin entropy increases, almost linearly.
  \item When $A_1 < A < A_2$ the basin entropy decreases rapidly.
  \item When $A_2 < A < A_3$ the basin entropy increases.
  \item When $A_3 < A < A_4$ the basin entropy decreases.
  \item When $A_4 < A < 1$ the basin entropy increases, while for $A > 1$ the tendency is reversed.
\end{itemize}
It is interesting to note that the lowest values of the basin entropy are observed exactly at the four critical values of the oblateness coefficient. This phenomenon can be explained if we take into account that for these values of $A$ the total number of roots decreases from five to three (when $A = A_1$ or $A = A_3$), two (when $A = A_2$) or even one (when $A = A_4$). Therefore the mixture of the several types of basins on the complex plane (even when the initial conditions which lead to infinity are counted as an additional type of basin) becomes leaner, which implies that the degree of fractality is reduced.

\section{Concluding remarks}
\label{conc}

The Newton-Raphson basins of convergence were numerically explored in the Sitnikov four-body problem, with non-spherical primaries. In particular, we demonstrated how the oblateness coefficient $A$ influences the position of the roots on the complex plane. The Newton-Raphson optimal iterative scheme was used for revealing the corresponding basins of convergence on the complex plane. These convergence domains play a significant role, since they explain how each point of the complex plane is numerically attracted by the equilibrium points of the system, which act, in a way, as attractors. We managed to monitor how the Newton-Raphson basins of convergence evolve as a function of the oblateness coefficient. Another important aspect of this work was the relation between the basins of convergence and the corresponding number of required iterations and the respective probability distributions.

As far as we know, this is the first time that the Newton-Raphson basins of convergence in the Sitnikov four-body problem are numerically investigated in such a systematic and thorough manner. On this basis, the presented results are novel and this is exactly the contribution of the present work.

The most important conclusions of our numerical analysis are summarized in the following list:
\begin{enumerate}
  \item Real and imaginary roots are only possible when the primaries are prolate $(A < 0)$. On the other hand, when the primary bodies have an oblate shape $(A > 0)$ the corresponding roots are always conjugate complex.
  \item It was found that all the basins of convergence, corresponding to all five roots, have finite area, regardless the particular value of the oblateness coefficient.
  \item Our numerical analysis indicates that the vast majority of the complex plane is covered by initial conditions which do not converge to any of the five roots. Furthermore, additional computations revealed that for all these initial conditions the Newton-Raphson iterator leads to extremely large complex numbers (either real or imaginary), which implies that these initial conditions tend asymptotically to infinity.
  \item Near the critical values of the oblateness coefficient we identified several types of converging areas for which the corresponding number of required iterations is relatively high, with respect to near by basins of other roots. We suspect that this phenomenon is inextricably linked with the fact that near these critical points the dynamics of the system, such as the total number of the equilibrium points (roots), changes.
  \item The highest values of the basin entropy, $S_b$, have been measured near the vicinity of the critical values of the oblateness coefficient, while the lowest values of $S_b$ were identified exactly at the critical values, where the total number of the roots of the system decreases.
\end{enumerate}

A double precision numerical code, written in standard \verb!FORTRAN 77! \cite{PTVF92}, was used for the classification of the initial conditions. In addition, for all the graphical illustration of the paper we used the latest version 11.2 of Mathematica$^{\circledR}$ \cite{W03}. Using an Intel$^{\circledR}$ Quad-Core\textsuperscript{TM} i7 2.4 GHz PC the required CPU time, for the classification of each set of initial conditions, was about 5 minutes.

In the future, it would be very interesting to use other types of iterative schemes and compare the similarities as well as the differences on the corresponding basins of convergence. More precisely, using iterative methods of higher order, with respect to the classical Newton-Raphson method of second order, would be an ideal starting point, for demystifying the secrets of this active field of research.

%\section*{Acknowledgments}

%I would like to express my warmest thanks to the anonymous referee for the careful reading of the manuscript and for all the apt suggestions and comments which allowed us to improve both the quality and the clarity of the paper.

%\end{multicols}

\end{document}